\documentclass{appolb}
\usepackage{epsfig}

\usepackage{graphicx}  
\usepackage{subfigure}

\usepackage{amsmath}

\begin{document}
\title{ Mass and decay of the $s\bar{s}$ member of the $1^{3}F_{4}$ meson nonet 
}
\author{Xue-Chao Feng $^{1}$\thanks{fxchao@zzuli.edu.cn}, Ke-Wei Wei $^{2}$\thanks{weikw@ihep.ac.cn}
\address{$^{1}$ College of Physics and Electronic Engineering, Zhengzhou University of Light Industry, 450002 Zhengzhou, China
\\ $^{2}$ School of Science, Henan University of Engineering, 451191 Zhengzhou, China } }

\maketitle
\begin{abstract}

The mass and decay of the $s\bar{s}$ member of the $1^{3}F_{4}$ meson nonet are investigated in the framework of the Regge phenomenology and the $^{3}P_{0}$ model. We propose, based on the results, that the assignment of the $s\bar{s}$  member of the $1^{3}F_{4}$ meson nonet will require additional testing in the future. Our results also provide information for future studies of the $1^{3}F_{4}$ meson nonet.

\end{abstract}

\section{Introduction}

In the 1970s, quantum chromodynamics was developed as a theory to describe strong interactions. With the discovery of asymptotic freedom in 1973, it was largely accepted since it satisfactorily explained some of the perplexing experimental results of the time. Nevertheless, there are still difficulties with the properties of low-energy QCD that require additional investigation, since the perturbation theory has been shown to be effective in the high-energy region, but is inapplicable to the low-energy scale. To investigate the properties of hadrons, such as mass and decay, various phenomenological models that embody the most essential properties of strong QCD have been constructed. The results derived from the theoretical model have dual applicability. On the one hand, it may be used to assess the validity of the model, and on the other, it can utilize these estimated properties to map the hadron spectrum. Inspired by this, we investigate the mass spectrum and decay of $1^{3}F_{4}$ nonet in this work. The investigation into the $1^{3}F_{4}$ nonet began more than 10 years ago. In this work, we believe it is vital to re-evaluate this issue because experimental data is continually updated throughout time.  In Table 1, we contrast PDG data from 2002, 2012, and 2022 editions. From Table 1, $a_{4}(1970)$, $f_{4}(2050)$, and $K^{\ast}_{4}(2045)$
\begin{center}
\indent\\ \footnotesize Table 1. Masses and decay widths of $J^{PC}=4^{++}$ meson states (in units of MeV).  $^{\dag}$ The states listed as ``further states'' in the PDG \cite{ParticleDataGroup:2022pth}.
\label{Tab:t1}
\begin{tabular}{llllllllll}
\\ \hline\hline
 PDG               & State       &Mass               & Width
\\
\hline
 $2022         $ \cite{ParticleDataGroup:2022pth}   &$a_{4}(1970)$             &  $1967\pm 16$             & $334^{+15}_{-18}$          &
\\
                   &$f_{4}(2050)$             &  $2018\pm 11$             & $237\pm 18$          &
\\
                   &$K^{\ast}_{4}(2045)$      &  $2048^{+8}_{-9}$         & $199^{+27}_{-19}$          &
\\
                   &$f_{4}(2300)^{\dag}$      &  $2320\pm 60$             & $250\pm 80$          &
\\
                   &$X(2000)^{\dag}$         &  $1998\pm 3\pm 5$         & $<15$           &
\\
                   &$a_{4}(2255)^{\dag}$      &  $2237 \pm 5$            & $291\pm 12$       &
\\ \hline
 $2012         $\cite{ParticleDataGroup:2012pjm}   &$a_{4}(2040)$             &  $1996^{+10}_{-9}$        & $255^{+28}_{-24}$          &
\\
                   &$f_{4}(2050)$             &  $2018\pm 11$             & $237\pm 18$          &
\\
                   &$K^{\ast}_{4}(2045)$      &  $2045\pm 9$             & $198\pm 30$          &
\\
                   &$f_{J}(2220)^{\dag}$      &  $2231.1\pm 3.5$          & $23^{+8}_{-7}$          &
\\
                   &$f_{4}(2300)^{\dag}$      &  $2320\pm 60$             & $250\pm 80$          &
\\
                   &$X(2000)^{\dag}$         &  $1998\pm 3\pm 5$         & $<15$           &
\\ \hline
 $2002         $ \cite{ParticleDataGroup:2002ivw}   &$a_{4}(2040)$             &  $2011\pm 13$             & $360\pm 40$          &
\\
                   &$f_{4}(2050)$             &  $2025\pm 8$             & $194\pm 13$          &
\\
                   &$K^{\ast}_{4}(2045)$      &  $2045\pm 8$         & $198\pm 30$          &
\\
                   &$f_{J}(2200)^{\dag}$      &  $2231.1\pm 3.5$          & $23^{+8}_{-7}$          &
\\
                   &$f_{4}(2300)^{\dag}$      &  $2332\pm 15$             & $260\pm 57$          &

\\ \hline

\end{tabular}
\end{center}
 are well established as members of the $1^{3}F_{4}$ meson nonet, despite the fact that the mass of $a_{4}(1970)$ has decreased by approximately 50 MeV in 20 years. The $a_{4}(2225)$ was listed as ``further states'' and reported in $\bar{p}p\rightarrow \pi^{0}\eta$, $3\pi^{0}$, $\pi^{0}\eta'$ \cite{Anisovich:2001pn}, and $\pi^{-}p\rightarrow \eta\eta\pi^{0}$ \cite{Uman:2006xb}. Compared to other states, the $s\bar{s}$ member of $1^{3}F_{4}$ have not been clearly established yet. In previous years, the $f_{J}(2220)$ has been thoroughly investigated as a candidate for  $s\bar{s}$ member. However, considering that its width is only 23 MeV, it is too small for traditional mesons. Over the years, people have also made other speculations, such as Higgs boson \cite{Willey:1983kh}, bound state of colored scalars \cite{Shatz:1984nt}, four quark state \cite{Chao:1984bq,Pakvasa:1984wb}, $\Lambda\bar{\Lambda}$ bound state \cite{Ono:1986rj}, hybrid or glueball state \cite{Chanowitz:1983sd,Huang:1995td}. The $f_{4}(2300)$ also has $J^{PC}=4^{++}$ quantum number, although it is omitted from the summery table in the PDG \cite{ParticleDataGroup:2022pth}. The $f_{4}(2300)$ was first discovered in Ref. \cite{Carter:1978ux} and was subsequently observed in the reactions, $\bar{p}p\rightarrow K^{-}K^{+}$ \cite{Carter:1978ux},$\eta\pi^{0}\pi^{0}$ \cite{Anisovich:2000ut},$\pi\pi$ \cite{Dulude:1978kt,Hasan:1994he}, and $\pi^{-}p\rightarrow K^{+}K^{-}$ \cite{VES:2000xjq}. There are some theoretical studies on the properties of the observed $f_{4}(2300)$ state. In Ref. \cite{Masjuan:2012gc}, Masjuan et al. show that $f_{4}(2300)$ may be a $s\bar{s}$ member of $1^{3}F_{4}$. The paper by Pang et al. shows that $f_{4}(2300)$ may be the first radial state of $f_{4}(2050)$ \cite{Pang:2015eha}. In a word, the understanding of $s\bar{s}$ member of the $1^{3}F_{4}$ meson nonet is still unclear. In this work, considering the present research situation, we will systematically analyze the mass and decays of $s\bar{s}$ member of the $1^{3}F_{4}$ meson nonet.

This paper is organized as follows. In Section 2, a brief summary of the Regge phenomenology and the $^{3}P_{0}$  model will be presented. The numerical results of the $1^{3}F_{4}$ meson nonet are shown in Section 3, while the results are described in Section 4.

\section{Theoretical models}

\subsection{Regge phenomenology }

In this section, we will review the Regge phenomenology theory. In the 1960s, the Regge theory was developed, which connects the high-energy behavior of the scattering amplitude with singularities in the complex angular momentum plane of the partial wave amplitudes \cite{Regge:1959mz}. According to Regge theory, mesons have poles that shift in the plane of complex angular momentum as a function of their energy. Regge trajectories of hadrons are often shown on the $(J, M^{2})$ plane, and these plots are known as Chew-Frautschi plots (where $J$ and $M$ are the total spins and masses of the hadrons, respectively)\cite{Chew:1962eu}. Recent years have seen a resurgence in interest in the Regge theory due to the fact that it may be applied to the prediction of meson masses as well as the determination of the quantum numbers of newly detected states in experiments \cite{Masjuan:2012gc,Brisudova:1999ut,Anisovich:2000kxa,Chen:2018nnr,Chen:2021kfw}.
In the past two decades, the quasilinear Regge trajectory was used for studying hadron spectra and result in reasonable description for the hadron spectroscopy \cite{Anisovich:2000kxa,Guo:2008he,Li:2007px,Li:2004gu}.

Based on the assumption that the hadrons with identical $J^{PC}$ quantum numbers obey quasi-linear form of Regge trajectories, one have the following relations

\begin{equation}
\label{Eq1}
J=\alpha_{n\bar{n}}(0)+\alpha'_{n\bar{n}}M^2_{n\bar{n}},
\end{equation}
\begin{equation}
\label{Eq2}
J=\alpha_{n\bar{s}}(0)+\alpha'_{n\bar{s}}M^2_{n\bar{s}},
\end{equation}
\begin{equation}
\label{Eq3}
J=\alpha_{s\bar{s}}(0)+\alpha'_{s\bar{s}}M^2_{s\bar{s}},
\end{equation}
The $\alpha'$ and $\alpha$ are the slope and intercept of the Regge trajectory, respectively. In this work, the intercept and slope can be expressed as
\begin{equation}
\label{Eq4}
\alpha_{n\bar{n}}(0)+\alpha_{s\bar{s}}(0)=2\alpha_{n\bar{s}}(0),
\end{equation}
\begin{equation}
\label{Eq5}
\frac{1}{\alpha'_{n\bar{n}}}+\frac{1}{\alpha'_{s\bar{s}}}=\frac{2}{\alpha'_{n\bar{s}}}.
\end{equation}

The intercept relation was derived from the dual-resonance model \cite{Kawarabayashi:1969yd}, and is satisfied in two-dimensional QCD \cite{Brower:1977as}, the dual-analytic model \cite{Kobylinsky:1978db}, the quark bremsstrahlung model \cite{Dixit:1979mz}. The slope relation (\ref{Eq5}) was obtained in the framework of topological expansion and the $q\bar{q}$-string picture of hadrons \cite{Kaidalov:1980bq}.

From relations (\ref{Eq1})-(\ref{Eq5}), one has
\begin{equation}
\label{Eq6}
M^{2}_{n\bar{n}}\alpha'_{n\bar{n}}+M^{2}_{s\bar{s}}\alpha'_{s\bar{s}}=2M^{2}_{n\bar{s}}\alpha'_{n\bar{s}}.
\end{equation}

Taking into account the fact that it is assumed that the partners' trajectories would always have parallel slopes,\cite{Brisudova:1999ut,Li:2004gu}, that is to say,
\begin{equation}
  \label{Eq7}
  \left\{
   \begin{array}{c}
\alpha'_{n\bar{n}}(1^{3}P_{2})=\alpha'_{n\bar{n}}(1^{3}F_{4}) \\ \\
\alpha'_{n\bar{s}}(1^{3}P_{2})=\alpha'_{n\bar{s}}(1^{3}F_{4}) \\ \\
\alpha'_{s\bar{s}}(1^{3}P_{2})=\alpha'_{s\bar{s}}(1^{3}F_{4}) \\
    \end{array}
  \right. .
\end{equation}

The following relations, which are connected to meson multiplets with the same regge slopes, can be obtained by eliminating the slopes of the equations:

$$
\frac  { 4M^{2}_{n\bar{s}(1^{3}P_{2})}  M^{2}_{n\bar{n}(1^{3}F_{4})}-4M^{2}_{n\bar{n}(1^{3}P_{2})}   M^{2}_{n\bar{s}(1^{3}F_{4})}  }
       { M^{2}_{n\bar{n}(1^{3}P_{2})}   M^{2}_{s\bar{s}(1^{3}F_{4})}- M^{2}_{s\bar{s}(1^{3}P_{2})}   M^{2}_{n\bar{n}(1^{3}F_{4})}  }
$$
\begin{equation}
\label{Eq8}
= \frac { M^{2}_{n\bar{s}(1^{3}P_{2})} ( M^{2}_{n\bar{n}(1^{3}F_{4})} - M^{2}_{s\bar{s}(1^{3}F_{4})} )  -
          M^{2}_{n\bar{s}(1^{3}F_{4})} ( M^{2}_{n\bar{n}(1^{3}P_{2})} - M^{2}_{s\bar{s}(1^{3}P_{2})} )         }
        { M^{2}_{n\bar{s}(1^{3}P_{2})}   M^{2}_{s\bar{s}(1^{3}F_{4})}- M^{2}_{s\bar{s}(1^{3}P_{2})}   M^{2}_{n\bar{s}(1^{3}F_{4})}   },
\end{equation}
where $M_{n\bar{n}}$,  $M_{n\bar{s}}$, and  $M_{s\bar{s}}$ denote the mass corresponding to the meson nonet member.

\subsection{$^{3}P_{0}$ model}

The $^{3}P_{0}$ model, also known as the quark-pair creation (QPC) model, was first proposed by Micu, and developed in the 1970s by LeYaouanc et al \cite{Micu:1968mk,LeYaouanc:1972vsx,LeYaouanc:1974cvx,LeYaouanc:1977fsz}. To this day, the model is widely used to calculate the decay amplitude and decay branch ratios of hadrons and has achieved very good results \cite{Lu:2016bbk,Roberts:1992esl,Barnes:1996ff,Ackleh:1996yt,Close:2005se,Barnes:2005pb,Zhang:2006yj,Lu:2014zua}. In this model, the decay process $A\rightarrow BC$ occurs when the quark-antiquark pair produces a state suitable for quark rearrangement in the vacuum.  The transition operator $T$ of the decay $A \rightarrow BC $ in the $^{3}P_{0}$
model is denoted by

$$T=-3 \gamma \sum_m\langle 1 m 1-m \mid 00\rangle \int d^3 \boldsymbol{p}_3 d^3 \boldsymbol{p}_4 \delta^3\left(\boldsymbol{p}_3+\boldsymbol{p}_4\right)$$

\begin{equation}
\label{Eq9}
\times \mathcal{Y}_1^m\left(\frac{\boldsymbol{p}_3-\boldsymbol{p}_4}{2}\right) \chi_{1,-m}^{34} \phi_0^{34} \omega_0^{34} b_3^{\dagger}\left(\boldsymbol{p}_3\right) d_4^{\dagger}\left(\boldsymbol{p}_4\right)
\end{equation}
where $\boldsymbol{p}_3$ and $\boldsymbol{p}_4$ are the momentum of the created quark (antiquark). The dimensionless parameter $\gamma$ represents the strength of the quark-antiquark pair created from
the vacuum. $\chi_{1,-m}^{34}, \phi_0^{34}$, and $\omega_0^{34}$ are spin, flavor, and
color wave functions of the created quark-antiquark pair,
respectively. The partial wave amplitude $\mathcal{M}^{L S}(\boldsymbol{P})$ for the decay $A \rightarrow B+C$ may be written as\cite{Jacob:1959at}

$$
\mathcal{M}^{L S}(\boldsymbol{P})=\sum_{\substack{M_{J_B, M}, M_{J_C} \\ M_S M_L}}\left\langle L M_L S M_S \mid J_A M_{J_A}\right\rangle\left\langle J_B M_{J_B} J_C M_{J_C} \mid S M_S\right\rangle
$$

\begin{equation}
\label{Eq10}
\int d \Omega Y_{L M_L}^* \mathcal{M}^{M_{J_A} M_{J_B} M_{J_C}(\boldsymbol{P})}
\end{equation}

With the transition operator $T$, the helicity amplitude $\mathcal{M}^{M_{J_A} M_{J_B} M_{J_C}}(\boldsymbol{P})$ can be written as
\begin{equation}
\label{Eq11}
\langle B C|T| A\rangle=\delta^3\left(\boldsymbol{P}_A-\boldsymbol{P}_B-\boldsymbol{P}_C\right) \mathcal{M}^{M_{J_A} M_{J_B} M_{J_C}}(\boldsymbol{P})
\end{equation}
The detailed analysis process can be referred to in Refs. \cite{Li:2010vx,Li:2008mza}.

In Refs. \cite{Ackleh:1996yt}, Ackleh et al. developed a diagrammatic, momentum space formulation of the $^{3}P_{0}$ model to evaluate the partial width $\Gamma_{A \rightarrow B C}$

\begin{equation}
\label{Eq12}
\Gamma_{A \rightarrow B C}=2 \pi \frac{P E_{B} E_{C}}{M_{A}} \sum_{L S}\left( M_{LS}\right)^{2}
\end{equation}
where $P$ is the decay momentum, $E_{B}$ and $E_{C}$ are the energies of mesons $B$ and $C$, in the rest frame of A,

$$
P=\frac{\left[\left(M_{A}^{2}-\left(M_{B}+M_{C}\right)^{2}\right)\left(M_{A}^{2}-\left(M_{B}-M_{C}\right)^{2}\right)\right]^{1 / 2}}{2 M_{A}}
$$

$$
E_{B}=\frac{M_{A}^{2}-M_{C}^{2}+M_{B}^{2}}{2 M_{A}}
$$

$$
E_{B}=\frac{M_{A}^{2}+M_{C}^{2}-M_{B}^{2}}{2 M_{A}}
$$

$M_{A}$, $M_{B}$, and $M_{C}$ are the masses of mesons $A$, $B$, and $C$. $M_{LS}$ are proportional to an overall Gaussian in $\frac{P}{\beta}$ times a channel-dependent polynomial $\xi_{L S}(\frac{P}{\beta})$,
$$
M_{LS}=\frac{\gamma}{\pi^{1 / 4} \beta^{1 / 2}} \xi_{L S}(\frac{P}{\beta}) e^{-P^{2} / 12 \beta^{2}}
$$

In this work, we take $\beta=0.4GeV$ and $\gamma=0.4$ as input, which is used in Refs.\cite{Barnes:1996ff,Ackleh:1996yt}.

\section{NUMERICAL RESULTS}

As presented in Eq. (\ref{Eq8}), the masses of $n\bar{n}$, $n\bar{s}$, $s\bar{s}$ of the $1^{3}P_{2}$ state are related to the masses of $n\bar{n}$, $n\bar{s}$, $s\bar{s}$ of the $1^{3}F_{4}$ state, inserting the values $M_{n\bar{n}(1^{3}P_{2})}$, $M_{n\bar{s}(1^{3}P_{2})}$, $M_{s\bar{s}(1^{3}P_{2})}$, $M_{n\bar{n}(1^{3}F_{4})}$ , and $M_{n\bar{s}(1^{3}F_{4})}$ , the mass of the member of $1^{3}F_{4}$ is determined to be 2129$\pm$20 MeV. In this work, the values $M_{n\bar{n}(1^{3}P_{2})}$ , $M_{n\bar{s}(1^{3}P_{2})}$ , $M_{n\bar{n}(1^{3}F_{4})}$, and $M_{n\bar{s}(1^{3}F_{4})}$ are taken from PDG, and the $M_{s\bar{s}(1^{3}P_{2})}$ is calculated using the average values in Table 2.
\begin{center}
\indent\\ \footnotesize Table 2 The mass of $s\bar{s}$ member of the $1^{3}P_{2}$ in different theoretical models (in units of MeV).
\label{Tab:t2}
\begin{tabular}{llllllllllll}
\hline \hline
&\cite{Brisudova:1999ut}& \cite{Li:2004gu}   &\cite{Godfrey:1985xj} & \cite{Ebert:2009ub}&\cite{Xiao:2019qhl}&\cite{Li:2020xzs}&\cite{Burakovsky:1999ug} &average

\\ \hline
$M_{1^{3}P_{2}(s\bar{s})}$
&1544                    & 1546                 &1530                  &1529                 &1539              &1513     & 1550           & 1536$\pm$11MeV

\\ \hline\hline
\end{tabular}
\end{center}

Based on the previously determined mass value, we used the $^{3}P_{0}$ model to investigate the decays of the $1^{3}F_{4}(s\bar{s})$ state. By inserting the decay polynomials $\xi_{L S}(\frac{P}{\beta})$ listed in Appendix A, we obtain the decay results from Eq. (\ref{Eq12}). The results are shown in Table 3.

\begin{center}
\indent\\ \footnotesize Table 3. Strong decay properties for the $1^{3}F_{4}(s\bar{s})$ state(in units of MeV). The initial
state mass is set to be 2129 $\pm$ 20 MeV, and the masses of all the final states are taken from PDG.
\label{Tab:t3}
\begin{tabular}{llllllllllll}

\hline
\hline
  Decay mode                                   &Present work & Decay mode                                 & Present work
\\
\hline
 $1^{3}F_{4}(s\bar{s})\rightarrow K K     $         & 25.9 $\pm$ 2.5   & $1^{3}F_{4}(s\bar{s})\rightarrow K K_(1460)   $          & 0
\\
 $1^{3}F_{4}(s\bar{s})\rightarrow K K^{*} $         & 17.9 $\pm$ 2.5     & $1^{3}F_{4}(s\bar{s})\rightarrow \eta \eta'   $          & 2.1 $\pm$ 0.4
\\
$1^{3}F_{4}(s\bar{s})\rightarrow K^{*}K^{*}$        & 37.6 $\pm$ 4.2     & $1^{3}F_{4}(s\bar{s})\rightarrow \eta' \eta'$          & 0
\\
$1^{3}F_{4}(s\bar{s})\rightarrow K K_{1}(1270)$     & 11 $\pm$ 2.2     & $1^{3}F_{4}(s\bar{s})\rightarrow \eta f_{1}(1420) $         & 0.2 $\pm$ 0.01
\\
 $1^{3}F_{4}(s\bar{s})\rightarrow K K_{1}(1400)$    & 0.6 $\pm$ 0.2    & $1^{3}F_{4}(s\bar{s})\rightarrow \eta f_{2}(1525) $      & 0
\\
$1^{3}F_{4}(s\bar{s})\rightarrow K K_{2}^{*}(1430) $ & 1.6 $\pm$ 0.6     & $1^{3}F_{4}(s\bar{s})\rightarrow \eta \eta(1475) $      & 0
\\
$1^{3}F_{4}(s\bar{s})\rightarrow K K^{*}(1414)$     & 0.2  $\pm$ 0.05   & $1^{3}F_{4}(s\bar{s})\rightarrow \phi\phi$              & 2.7 $\pm$ 1.5

\\
\hline
\hline
\end{tabular}

\end{center}


\section{Conclusion}

According to the introduction in Section 1, it is important to study the mass spectrum and decay of mesons using phenomenological models. In the present work, we investigated the mass spectrum and decay of the $s\bar{s}$ member of the $1^{3}F_{4}$ meson nonet. In the framework of Regge phenomenology, the mass of $s\bar{s}$ member is determined to be 2129$\pm$ 20 MeV, which is consistent with prediction with the covariant oscillator quark model \cite{Ishida:1986vn}, while it is about 100 MeV smaller than the prediction with relativized quark model \cite{Godfrey:1985xj,Xiao:2019qhl}. In the new edition of PDG, the $f_{4}(2300)$ has a definite quantum numbers $4^{++}$, although it is omitted from the summary table \cite{ParticleDataGroup:2022pth}. The Ref. \cite{Masjuan:2012gc} suggest that this state may be a $s\bar{s}$ member of $1^{3}F_{4}$ meson nonet. In our work, considering the fact that mass is about 200 MeV smaller than the measured mass of $f_{4}(2300)$, our analysis does not support this assignment. Apart from the mass spectrum, the decay was analyzed using the $^{3}P_{0}$ model, and results indicate that $K K$ and $K^{*}K^{*}$ are the two primary decay modes. Due to limited understanding of the $s\bar{s}$ member of the $1^{3}F_{4}$ meson nonet, we hope that these analyses can provide some meaningful guidance in understanding the $1^{3}F_{4}$ meson nonet.

\appendix
\section{the polynomial $\xi_{L S}(\frac{P}{\beta})$ for the $1^{3}F_{4}$ meson nonet decay in $^{3}P_{0}$ model.}
\begin{center}
\indent\\
\begin{tabular}{llllll}

\hline
\hline
  Decay mode                                   &$\xi_{L S}(\frac{P}{\beta})$
\\
\hline
 $1^{3}F_{4}(s\bar{s})\rightarrow K K     $         & $\xi_{{40}{(^{3}F_{4} \rightarrow ^{1}S_{0}+ ^{1}S_{0})}}= 4.29 \times 10^{-1} \qquad  ^{1}{G}_{4}$
\\
\\
 $1^{3}F_{4}(s\bar{s})\rightarrow K K^{*} $         & $\xi_{{41}{(^{3}F_{4} \rightarrow ^{1}S_{0}+ ^{3}S_{1})}}= -1.43 \times 10^{-1} \qquad ^{3}{G}_{4}$
 \\
\\
$1^{3}F_{4}(s\bar{s})\rightarrow K^{*}K^{*}$        & $\xi_{{LS}{(^{3}F_{4} \rightarrow ^{3}S_{1}+ ^{3}S_{1})}}=
 \left\{\begin{array}{cc}
   -4.87 \times 10^{-1}                    & \qquad{ }^{5} {D}_{4} \\
   3.61 \times 10^{-2}                     & \qquad{ }^{1} {G}_{4} \\
   -7.14 \times 10^{-1}                    & \qquad{ }^{5} {G}_{4}
\end{array}\right.
$
\\
\\

$1^{3}F_{4}(s\bar{s})\rightarrow K K_{1}(1270)$     & $\xi_{{LS}{(^{3}F_{4} \rightarrow ^{1}P_{1}+ ^{1}S_{0})}}=
 \left\{\begin{array}{cc}
  -1.98 \times 10^{-1}                    & \qquad{ }^{3} {F}_{4} \\
  -1.40 \times 10^{-1}                    & \qquad{ }^{3} {H}_{4} \\

\end{array}\right.
$
\\
\\
 $1^{3}F_{4}(s\bar{s})\rightarrow K K_{1}(1400)$    & $\xi_{{LS}{(^{3}F_{4} \rightarrow ^{3}P_{1}+ ^{1}S_{0})}}=
 \left\{\begin{array}{cc}
   5.11 \times 10^{-2}                   & \qquad{ }^{3} {F}_{4} \\
   2.65 \times 10^{-3}                   & \qquad{ }^{3} {H}_{4} \\

\end{array}\right.
$
\\
\\
$1^{3}F_{4}(s\bar{s})\rightarrow K K_{2}^{*}(1430) $ & $\xi_{{LS}{(^{3}F_{4} \rightarrow ^{3}P_{2}+ ^{1}S_{0})}}=
 \left\{\begin{array}{cc}
   8.53 \times 10^{-2}                   & \qquad{ }^{5} {F}_{4} \\
   2.38 \times 10^{-2}                   & \qquad{ }^{5} {H}_{4} \\

\end{array}\right.
$
\\
\\
$1^{3}F_{4}(s\bar{s})\rightarrow K K^{*}(1414)$     & $\xi_{{41}{(^{3}F_{4} \rightarrow ^{1}S_{0}+ 2^{3}S_{1})}}= -2.13 \times 10^{-2}     \qquad{ }^{3} {G}_{4}
$
\\
\\
 $1^{3}F_{4}(s\bar{s})\rightarrow K K_(1460)   $          & $\xi_{{40}{(^{3}F_{4} \rightarrow ^{1}S_{0}+ 2^{1}S_{0})}}= 1.90\times 10^{-2}  \qquad{ }^{1} {G}_{4}
$
\\
\\
$1^{3}F_{4}(s\bar{s})\rightarrow \eta \eta   $          & $\xi_{{40}{(^{3}F_{4} \rightarrow ^{1}S_{0}+ ^{1}S_{0})}}= 3.79 \times 10^{-1}  \qquad{ }^{1} {G}_{4}
$
\\
\\
$1^{3}F_{4}(s\bar{s})\rightarrow \eta \eta'   $          & $\xi_{{40}{(^{3}F_{4} \rightarrow ^{1}S_{0}+ ^{1}S_{0})}}= 1.63 \times 10^{-1}   \qquad{ }^{1} {G}_{4}
$

\\
\\
$1^{3}F_{4}(s\bar{s})\rightarrow \eta' \eta'   $          & $\xi_{{40}{(^{3}F_{4} \rightarrow ^{1}S_{0}+ ^{}S_{0})}}= 2.55 \times 10^{-2}   \qquad{ }^{1} {G}_{4}
$
\\
\\
$1^{3}F_{4}(s\bar{s})\rightarrow \eta f_{1}(1420) $         & $\xi_{{LS}{(^{3}F_{4} \rightarrow ^{1}S_{0}+ ^{3}P_{0})}}=
 \left\{\begin{array}{cc}
  4.31 \times 10^{-2}                   & \qquad{ }^{3} {F}_{4} \\
   -1.97 \times 10^{-3}                  & \qquad{ }^{3} {H}_{4} \\

\end{array}\right.
$

\\
\\
$1^{3}F_{4}(s\bar{s})\rightarrow \eta f_{2}(1525) $      & $\xi_{{LS}{(^{3}F_{4} \rightarrow ^{1}S_{0}+ ^{3}P_{2})}}=
 \left\{\begin{array}{cc}
   1.54 \times 10^{-2}                  & \qquad{ }^{5} {F}_{4} \\
   1.34 \times 10^{-4}                  & \qquad{ }^{5} {H}_{4} \\
\end{array}\right.
$

\\
\\
$1^{3}F_{4}(s\bar{s})\rightarrow \eta \eta(1475) $      & $\xi_{{40}{(^{3}F_{4} \rightarrow ^{1}S_{0}+ 2^{1}S_{0})}}=    4.69 \times 10^{-3} \qquad{ }^{1} {G}_{4}
$

\\
\\
$1^{3}F_{4}(s\bar{s})\rightarrow \phi\phi$              & $\xi_{{LS}{(^{3}F_{4} \rightarrow ^{3}S_{1}+ ^{3}S_{1})}}=
 \left\{\begin{array}{cc}
  -1.60 \times 10^{-1}                   & \qquad{ }^{5} {D}_{4} \\
   2.78 \times 10^{-3}                  & \qquad{ }^{1} {G}_{4} \\
   -5.51 \times 10^{-3}                  & \qquad{ }^{5} {G}_{4} \\
\end{array}\right.
$
\\

\hline
\hline
\end{tabular}

\end{center}


\end{document}